\documentclass[12pt]{article}
\usepackage{graphicx}
\usepackage{amsmath}
\usepackage{amssymb}
\usepackage{graphicx}
\usepackage[utf8]{inputenc}
\usepackage{amsmath}
\usepackage{amsfonts}
\usepackage{amssymb}
\usepackage{graphicx}
\usepackage{color}

\begin{document}

	\begin{center}
		\Large {\bf  Electromagnetic Waves  Generated \\ by Null Cosmic Strings Passing   Pulsars}
	\end{center}

	\bigskip
	\bigskip
	
	\begin{center}
		D.V. Fursaev$^\dag$ and I.G. Pirozhenko$^{\dag\ddag}$
	\end{center}
	
	\bigskip
	\bigskip
	\date{today}
	
	\begin{center}
		{\dag \it  Bogoliubov Laboratory of Theoretical Physics\\
			Joint Institute for Nuclear Research\\
			141 980, Dubna, Moscow Region, Russia}\\
			
			and\\
			
		\ddag	{\it Dubna State University,
			Universitetskaya st. 19\\ }
		\medskip
	\end{center}

\begin{abstract}
Null cosmic strings disturb electromagnetic (EM) fields of charged sources and sources with magnetic moments.
As has been recently shown by the authors, these perturbations result in a self-force acting on the sources and
create
EM waves outgoing from the sources. We develop an analytic approximation
for asymptotic of the EM waves at the future null infinity
and calculate radiation fluxes for sources of the both types. For magnetic-dipole-like sources the radiation flux depends on orientation
of the magnetic moment with respect to the string. Estimates show that the peak
power of the  radiation can be quite large for null strings moving near pulsars and considerably large in case of magnetars. The
string generated variations of the luminosities of the stars
can be used as a potential experimental signature of null cosmic strings.
\end{abstract}

\newpage

\section{Introduction}\label{intr}

As has been recently shown \cite{Fursaev:2022ayo}, null cosmic strings
disturb electric fields of charged sources
and produce electromagnetic (EM) pulses  outgoing from the sources. The aim of the present work, based on the approach developed in
\cite{Fursaev:2022ayo}, is two-fold. First, it is to describe in more detail properties of EM waves at future null infinity generated by straight null cosmic strings in a locally Minkowsky space-time. Second, it is to study analogous effects for sources with magnetic moments, such as pulsars and magnetars.
As opposed to electrically charged objects the case of pulsars and magnetars is more interesting from astrophysical
point of view. We show that variations of luminosities of such objects induced by the null strings can be large enough to be experimentally observable.

The null strings are one-dimensional objects whose points move  along  trajectories of light rays, orthogonally to strings \cite{Schild:1976vq}.   As a result, the null strings exhibit optical properties: they behave as one-dimensional null geodesic congruences  governed by an analogue of the Sachs' optical equation \cite {Fursaev:2021xlm}. The world-sheets of null strings may develop caustics  \cite{Davydov:2022qil}.

The null cosmic strings, like tensile
cosmic strings \cite{Kibble:1976sj}, \cite{Vilenkin:2000jqa}, are hypothetical astrophysical objects which might have been produced in the very early Universe,  at the Planckian epoch \cite{GM1}-\cite{Xu:2020nlh}.  Possible astrophysical and cosmological  effects  of null cosmic strings, such as deviations of light rays and trajectories of particles in the gravitational field of strings \cite{Fursaev:2017aap}, \cite{Fursaev:2018spa} as well as scattering
of strings by massive sources \cite{Davydov:2022qil}, are similar to those of the tensile strings.

Gravitational fields of null cosmic strings can be described in terms of holonomies around the string world-sheet.
The holonomies  are null rotations of the Lorentz group \cite{vandeMeent:2012gb}, \cite{Fursaev:2017aap}
with the group parameter determined by the string energy per unit  length. The null rotations
have fixed points on the string world-sheet ${\cal S}$. The world-sheet belongs to a null hypersurface ${\cal H}$ which is the string event horizon.

To find trajectories of particles and light rays near null cosmic strings one should set \cite{Fursaev:2017aap}, for matter crossing the string horizon, ``initial'' data on ${\cal H}$ to ensure the
required holonomy transformations.  Such an approach has been generalized in \cite{Fursaev:2022ayo} to describe
evolution of classical fields under the gravity of the null strings. Finding solutions to wave equations in
a free field theory on a space-time of a null string, in the domain above
${\cal H}$, is equivalent  to solving a characteristic problem with initial data on ${\cal H}$ determined by incoming data. The initial data are null rotated (or, better say, transformed by the Carroll group of ${\cal H}$) so that to ensure the required holonomy.  It is this approach we use in the present work to describe perturbations caused by null cosmic strings on EM fields of point-like sources.

In  \cite{Fursaev:2022ayo} we found a useful integral representation for perturbations of EM fields of point electric charges and straight strings.
We generalize this result to point magnetic-dipole-like sources (MD-sources) and develop an approximation for the perturbations at future null
infinity. The asymptotic for angular components
of the vector potential at large distances $r$ from the source look as
\begin{equation}\label{i.1}
	A_{B}(r,U,\Omega)\simeq a_B (U,\Omega) + b_B (\Omega)\ln r/\varrho+O(r^{-1}\ln r)~~~,
\end{equation}
where $U$ is a retarded time, vector fields $a_B$, $b_B$ are in the tangent space of $S^2$, and $\varrho$ is a  dimensional parameter related to the approximation.
The asymptotic holds if $r$ is large enough with respect to $U$ and with respect to the impact parameter between the string and the source.
Equation (\ref{i.1}) yields a finite energy flux
at large $r$
\begin{equation}\label{i.2}
\lim_{r\to \infty}	\partial_t E(r,t)=\int d\Omega ~\dot{a}_A \dot{a}^A~~,
\end{equation}
where $\dot{a}_A\equiv\partial_Ua_A$, and index $A$ is risen  with the help of the metric on $S^2$.
In the found approximation, we analytically calculate $\dot{a}_A \dot{a}^A$, the density of the flux,  for different sources
and study its properties.  The  logarithmic term in (\ref{i.1})  appears since the Sommerfeld radiation condition is violated in the presence of null strings. This term, however, does not affect the flux (\ref{i.2}).

The paper is organized as follows.   In Sec.  \ref{method} we introduce necessary notions and describe the method. In Sec.  \ref{sym}
we pay a special attention to the isometry group near the string world-sheet
$\cal S$. On the string horizon, the null rotations induce  the Carroll transformations
\cite{Duval:2014lpa},\cite{Duval:2014uoa},\cite{Ciambelli:2023tzb},\cite{Ciambelli:2023xqk}, which we use to set
initial data on $\cal H$.  The approach \cite{Fursaev:2017aap} and the characteristic problem
for EM fields \cite{Fursaev:2022ayo} are formulated in Sec. \ref{problem}. The problem is characteristic since $\cal H$ is a null hypersurface.
Point-like sources we consider are defined in Sec. \ref{definition}.
Fields of MD-sources may be viewed as a far-distance approximation of fields of finite size bodies
with magnetic moments.  Solutions for characteristic problem of vector fields can be formulated in terms
of solutions for auxiliary scalar problem defined in Sec. \ref{aux}. Asymptotic properties of EM waves and energy fluxes
generated from
electric and MD sources are considered in detail in Sections \ref{elrad} and \ref{marad}, respectively.
Here we provide explicit expressions for quantities in (\ref{i.1})  and (\ref{i.2}).
Possibilities to detect EM pulses generated by null cosmic strings near pulsars and magnetars are discussed
in Sec. \ref{exp}. A summary and discussion of our results can be found in Sec. \ref{sum}.
Approximations for integral representations of perturbations caused by null strings
are developed in Appendix \ref{App1}. Some technicalities related to perturbations for MD sources
are given in Appendix \ref{App2}.

\section{Description of the method}\label{method}
\subsection{Null rotations and Carroll transformations}\label{sym}
\setcounter{equation}0

We consider classical electrodynamics
\begin{equation}\label{1.1}
\partial_\mu F^{\mu\nu}=j^\nu~~,
\end{equation}
where $F_{\mu\nu}=\partial_\mu A_\nu - \partial_\nu A_\mu$. The current $j^\mu$, $\partial j=0$, will correspond to a point-like
source, either of electric or MD type.  We study solutions of (\ref{1.1}) near a straight cosmic string which is stretched along $z$-axis and moves along $x$-axis.
The space-time is locally Minkowsky $R^{1,3}$, and we use the light-cone coordinates  $v=t+x$, $u=t-x$, where the metric is
\begin{equation}\label{1.2}
ds^2=-dv du +dy^2+dz^2~~.
\end{equation}
The string world-sheet $\cal S$ can be defined by equations $u=y=0$.

A number of definitions are needed for further purposes.
We use the null rotations $x^\mu=M^\mu_{~\nu}\left(\lambda\right)\bar{x}^\nu$ which leave invariant (\ref{1.2}):
\begin{equation}\label{1.3}
u=\bar{u}~~,~~
v=\bar{v}+2\lambda \bar{y}+\lambda^2 \bar{u}~~,~~
y=\bar{y}+\lambda \bar{u}~~,~~z=\bar{z}~~,
\end{equation}
where $\lambda$ is some real parameter.  Transformations of quantities with lower indices are
\begin{equation}\label{1.4}
V_u=\bar{V}_u-\lambda \bar{V}_y+\lambda^2 \bar{V}_v~~,~~
V_v=\bar{V}_v~~,~~
V_y=\bar{V}_y-2\lambda \bar{V}_v~~,~~V_z=\bar{V}_z~~,
\end{equation}
or  $V_\mu=M_{\mu}^{~\nu}(\lambda )\bar{V}_\nu$, where $M_{\mu}^{~~\nu}=\eta_{\mu\mu'}\eta^{\nu\nu'}M^{\mu'}_{~~\nu'}$.
The null rotations make a parabolic subgroup of the Lorentz group.

For a null string with the world-sheet $u=y=0$ a parallel transport of a vector $V$ along a closed contour around the string
results in a null rotation, $V'=M(\omega)V$ with $\omega$ defined as, see \cite{vandeMeent:2012gb},
\begin{equation}\label{1.5}
\omega \equiv 8\pi GE~~.
\end{equation}
The world-sheet is a fixed point set of (\ref{1.3}).

The null hypersurface $u=0$ is the event horizon of the string. We denote  it
by ${\cal H}$. The properties
of null hypersurfaces are reviewed, for example, in \cite{Gourgoulhon:2005ng}. We enumerate coordinates $v,y,z$ on $\cal H$
by indices $a,b,..$. Coordinate transformations (\ref{1.3})  at $u=0$ (with $\lambda=\omega$) induce the change of coordinates
on $\cal H$:
\begin{equation}\label{1.14}
x^a=C^{a}_{~b}\bar{x}^b~~~\mbox{or}~~~{\bf x}=\bar{\bf x}+2\omega y{\bf{q}}~~,~~q^a=\delta^a_v~~,
\end{equation}
where ${\bf x}\equiv \{v,y,z\}$.  Matrices  $C^{a}_{~b}$ can be defined as
\begin{equation}\label{1.14b}
C^{a}_{~b}=M^{a}_{~b}(\omega)~~.
\end{equation}
Transformations (\ref{1.14}), (\ref{1.14b})  make the Carroll group of symmetries of the string horizon.
An introduction to the Carroll transformations can be found in \cite{Duval:2014lpa},\cite{Duval:2014uoa}.
The components of one-forms $\theta=\theta_a dx^a$ and vector fields $V=V^a\partial_a$ on $\cal H$, in a coordinate
basis, change as
\begin{equation}\label{1.18a}
\theta_a({\bf x})=C_{a}^{~~b}~\bar{\theta}_b(\bar{\bf x})~~,~~V^a({\bf x})=C^{a}_{~b}~\bar{V}^b(\bar{\bf x})~~,
\end{equation}
where $C_{a}^{~~b}=M_{a}^{~~b}$.
Since $M_b^{~u}=0$ the matrix  $C_{a}^{~~b}$ is inverse of $C^{a}_{~c}$.
Forms like $\theta(V)=\theta_a V^a$ are invariant with respect to
(\ref{1.18a}).  It should be noted, however, that indices $a,b$ cannot be risen or lowered since the metric of $\cal H$ is degenerate.

By following the method suggested in \cite{Fursaev:2017aap} one can construct a string space-time, which is locally flat but has the required holonomy  on $\cal S$.
One starts with $R^{1,3}$ which
 is decomposed onto two parts: $u<0$, and $u>0$.  Trajectories of particles and light rays at $u<0$ and $u>0$
can be called ingoing and outgoing trajectories, respectively.
To describe outgoing trajectories, one introduces two types
of coordinate charts: $R$- and $L$-charts, with cuts on the horizon either on the left
($u=0, y<0$) or on the right ($u=0, y>0$). The initial data on the string horizon are related to the ingoing data via null rotations (\ref{1.3}) taken at $u=0$. For  brevity the right  ($u=0, y>0$) and the left ($u=0, y<0$) parts of $\cal H$ will be denoted as
${\cal H}_+$
and ${\cal H}_-$.

For the $R$-charts the cut is along ${\cal H}_-$. If $x^\mu$ and $\bar{x}^\mu$ are, respectively, the coordinates above and
below the horizon the transition conditions on $\cal H$ in the $R$-charts look as
\begin{equation}\label{1.6}
x^\mu=\bar{x}^\mu\mid_{{\cal H}_+}~~\mbox{or}~~x^a=\bar{x}^a~~,
\end{equation}
\begin{equation}\label{1.7}
x^\mu=M^\mu_{~\nu}\left(\omega\right)\bar{x}^\nu\mid_{{\cal H}_-}~~\mbox{or}~~x^a=C^a_{~b}\bar{x}^b~~.
\end{equation}
Analogously, 3 components of 4-velocities of particles and light rays (with the lower indices)
change on ${\cal H}_-$ as $u_a=C_a^{~b}\bar{u}_b$.  Change of the $u_u$-component
can be found by requiring the invariance of $u^\mu u_\mu$, which implies that $u^\mu$ experiences
the 4-dimensional null rotation on ${\cal H}_-$.

Coordinate transformations (\ref{1.7}) are reduced to a linear supertranslation
\begin{equation}\label{1.8}
v=\bar{v}+2\omega y~~,~~y<0~~.
\end{equation}
Hence the gravitational field of a null string is a particular example of a gravitational shockwave background.
The method we follow is close to a more general approach suggested by Penrose  \cite{Penrose:1972xrn}.

As a result of the Lorenz invariance of the theory the descriptions based on $R$- or $L$-charts are equivalent.
Without loss of the generality from now on we work with the $R$-charts.
In this case all field variables experience Carroll transformations on  ${\cal H}_-$,
and one needs to solve field equations in the domain $u>0$ with the initial data changed on ${\cal H}_-$.

\subsection{Characteristic problem for EM fields}\label{problem}

Consider the Maxwell theory on the string space-time. Let $\bar{A}_\mu$ be a solution to the problem
in the region $u<0$
\begin{equation}\label{1.1b}
\partial_\mu \bar{F}^{\mu\nu}=\bar{j}^\nu~~.
\end{equation}
The currents at $u>0$ and $u<0$ are related on $\cal H$ as
\begin{equation}\label{1.17}
j^\mu({\bf x})\mid_{{\cal H}_+}=\bar{j}^\mu({\bf x})~~,~~
j^\mu({\bf x})\mid_{{\cal H}_-}=M^{\mu}_{~\nu}(\omega)\bar{j}^\nu(\bar{\bf x})~~.
\end{equation}

We need to solve (\ref{1.1}) at $u>0$  by setting the following initial conditions at $u=0$:
\begin{equation}\label{1.11}
A_b (x)\mid_{\cal H}=a_b ({\bf x})~~,~~\bar{A}_b (x)\mid_{{\cal H}}=\bar{a}_b ({\bf x})
\end{equation}
\begin{equation}\label{1.12}
a_{b}({\bf x})=\bar{a}_b({\bf x})\mid_{{\cal H}_+}~~,~~
a_{b}({\bf x})=C_{b}^{~c}(\omega)\bar{a}_c(\bar{\bf x})\mid_{{\cal H}_-}~~.
\end{equation}

In the theory of hyperbolic second order  partial differential equations (PDE) a problem is called  the characteristic initial value problem \cite{Morse:1953} if the initial data are set on a null hypersurface. In this case the number of initial data are twice less,
and setting just $a_b$ is enough to determine  the solution (see discussion in \cite{Fursaev:2022ayo}).

The variation of the action on the string space-time, if we take into account (\ref{1.1}), (\ref{1.1b}), has the form
\begin{equation}\label{1.9}
\delta_A\left(-\frac 14  \int \sqrt{g}d^4x ~F^{\mu\nu}F_{\mu\nu}+I[A]\right)=
\frac 12
\int_{{\cal H}}d^3{\bf x} \left(\pi^b({\bf x}) \delta a_b ({\bf x})-\bar{\pi}^b ({\bf x})\delta \bar{a}_b ({\bf x})\right)~~,
\end{equation}
\begin{equation}\label{1.10}
\pi^b=F^{ub}~~,~~\bar{\pi}^b=\bar{F}^{ub}~~,~~b=v,y,z~~.
\end{equation}
Here $I[A]$ is the action of a charged matter, and $\pi^b$, $\bar{\pi}^b$ can be interpreted as canonical momenta conjugated with $a_b$, $\bar{a}_b$.

This variation vanishes given conditions (\ref{1.12}) since the momenta transform under the Carroll group as components
of a 3-vector,
\begin{equation}\label{1.16}
\pi^{b}({\bf x})=\bar{\pi}^b({\bf x})\mid_{{\cal H}_+}~~,~~
\pi^{b}({\bf x})=C^{b}_{~c}\bar{\pi}^c(\bar{\bf x})\mid_{{\cal H}_-}~~.
\end{equation}
This transformation law follows, for example, from the constraints
\begin{equation}\label{1.15}
\partial_b \pi^b=-j^u\mid_{\cal H}~~, ~~\partial_b \bar{\pi}^b=-\bar{j}^u\mid_{\cal H}~~,
\end{equation}
and the fact that $j^u=\bar{j}^u$.  The given arguments, of course, mean that the Maxwell strength is continuous across  $\cal H$ on the string space-time.

In what follows we suppose that charged particles do not cross the left part of the horizon, that is
$j^\mu=\bar{j}^\mu$ on ${\cal H}$.
One can
write a solution to (\ref{1.1}),  (\ref{1.11}), (\ref{1.12}) in the form:
\begin{equation}\label{1.18}
A_\mu(x)=\bar{A}_{\mu}(x)+A_{S,\mu}(x)~~.
\end{equation}
Here $\bar{A}_{\mu}$ is an extension of the solution  (\ref{1.1b}) to the region $u>0$, and
$A_{S,\mu}$ is a solution to a homogeneous characteristic initial value problem
\begin{equation}\label{1.19}
\partial_\mu F_S^{\mu\nu}=0~~,~~A_{S,b} (x)\mid_{{\cal H}}=a_{S,b}({\bf x})~~,
\end{equation}
\begin{equation}\label{1.20}
a_{S,b}({\bf x})\mid_{{\cal H}_+}=0~~,~~
a_{S,b}({\bf x})\mid_{{\cal H}_-}=C_{b}^{~c}(\omega)\bar{a}_c(\bar{\bf x})-\bar{a}_b({\bf x})
~~,
\end{equation}
$F_{S,\mu\nu}=\partial_\mu A_{S,\nu} - \partial_\nu A_{S,\mu}$.
One can interpret  $\bar{A}$ as a solution in the absence of the string, while
$A_{S}$ can be considered as a perturbation caused by the string. This perturbation:

i)  vanishes in the limit $\omega\to 0$, where $\omega$ is the energy of the string;

ii) depends on the choice of the source $j_\mu$  through initial data (\ref{1.20});

ii) can be written as
\begin{equation}\label{1.21}
A_{S,\mu}(x)={\cal A}^\omega_{\mu}(x)- {\cal A}_{\mu}(x)
~~,
\end{equation}
\begin{equation}\label{1.22}
{\cal A}^\omega_{\mu}(x)=M_\mu^{~\nu} (\omega) {\cal A}_\nu (\bar{x})~~,~~x^\mu=M^\mu_{~\nu}\left(\omega\right)\bar{x}^\nu
~~,
\end{equation}
where ${\cal A}_{\mu}(x)$ is a solution to homogeneous problem $\partial_\mu {\cal F}^{\mu\nu}=0$ with
the following initial data on $\cal  H$
\begin{equation}\label{1.23}
{\cal A}_{b}({\bf x})\mid_{{\cal H}_+}=0~~,~~{\cal A}_{b}({\bf x})\mid_{{\cal H}_-}=\bar{a}_b({\bf x})
~~.
\end{equation}
A geometrical interpretation of the perturbation is that at small $\omega$  it is the Lie derivative,
\begin{equation}\label{1.24}
A_S=-\omega {\cal L}_\zeta {\cal A}~~,
\end{equation}
generated by vector $\zeta^\mu=2y\delta^\mu_v+u\delta^\mu_y$ associated to null rotations (\ref{1.3}).

For the subsequent analysis, it is convenient to use the Lorentz gauge condition $\partial A=0$  since it is invariant under
null rotations
and can be imposed globally on the cosmic string space-time. Thus, equations for the perturbations caused by the string are reduced to
the problem:
\begin{equation}\label{1.25}
\Box A_{S,a}=0~~,~~A_{S,a} (x)\mid_{{\cal H}}=a_{S,a}({\bf x})~~
\end{equation}
with initial data (\ref{1.20}). Component $A_u$ is determined by condition $\partial A_S=0$.
The rest of the paper is devoted to studying solutions of (\ref{1.25}) for  different sources.

\section{Asymptotics and radiation energy flux}\label{Asympt}
\setcounter{equation}0

\subsection{Definitions}\label{definition}

\begin{figure}
	\begin{center}
	\includegraphics[width=6cm]{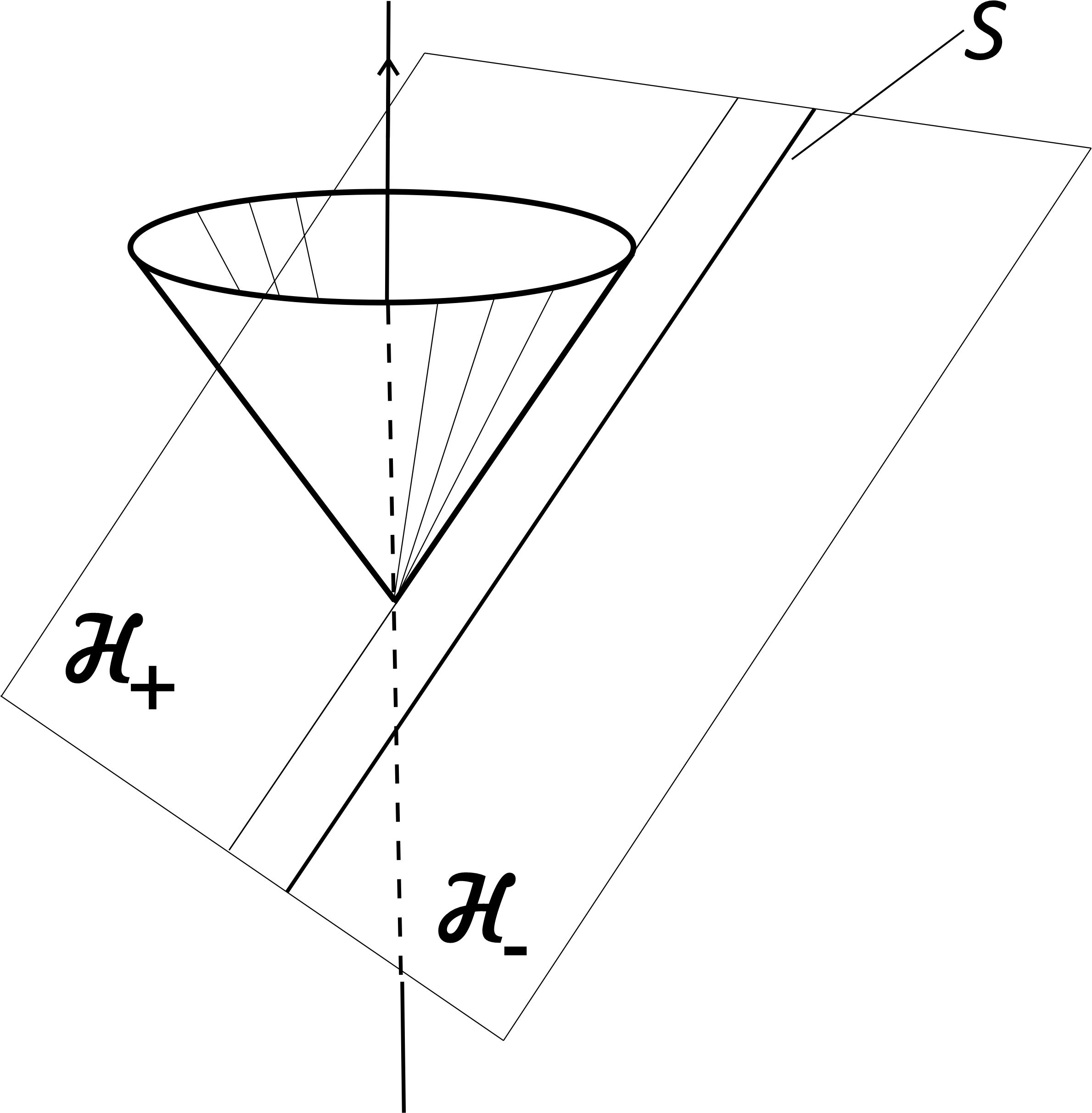}
	\end{center}
	\caption{shows the world-sheet of the string  $\cal S$, the string horizon ${\cal H}={\cal H}_+\cup {\cal H}_-$ and
a trajectory of point-like source (the dashed line) that crosses  ${\cal H}_+$.  At late times perturbation of EM field of the source caused by the string propagates along a null cone which is tangent to $\cal H$ and has the apex approximately at a point where the trajectory crosses the horizon.}
\label{Cone}
\end{figure}

Perturbations  $A_{S,\mu}$ will be considered for point-like sources of electric and magnetic types.
Without loss of the generality we suppose that the source is at rest at a point with coordinates $x_o=z_o=0,y_o=a>0$.
Since the string trajectory is $x=t$, $y=0$, we interpret $a$ as an impact parameter between the string and the source.
As we agreed, the source crosses ${\cal H}_+$, see Fig.~\ref{Cone}.

The sources of the electric type are just electric charges. The corresponding current  is
\begin{equation}\label{4.1}
	j_0=e\delta^{(3)}(\vec{x}-\vec{x}_o)~~~,~~~j_i(x)=0~~,~~ i=1,2,3~~.
\end{equation}
The field in the absence of the string is
\begin{equation}\label{4.2}
	\bar{A}_0(x)=e\phi(x)~~~,~~~\bar{A}_i(x)=0~~,
\end{equation}
\begin{equation}\label{coul}
	\phi(x,y,z)=-\frac{1}{4\pi}\frac{1}{\sqrt{x^2+(y-a)^2+z^2}}~~~.
\end{equation}
The sources of the MD type are described by the current
\begin{equation}\label{4.3}	
	j_0=0, \quad  j_i(x)=\varepsilon_{ijk} M_j \, \partial_k \delta^{(3)}(\vec{x}-\vec{x}_o)~~.
\end{equation}
where $M_i$ is a magnetic moment. The corresponding field in the absence of the string is
\begin{eqnarray}\label{4.4}
	\bar{A}_0(x)=0, \quad  \bar{A}_i(x)=\varepsilon_{ijk} M_j \partial_k \phi~~~,
\end{eqnarray}
which is the field of a magnetic dipole \cite{Landau2}.
We consider (\ref{4.4}) as an approximation for a EM field of a body with a magnetic moment.
Under certain assumptions (\ref{4.4}) can be used to describe the EM field of such objects as pulsars. Note that $M_i$ in (\ref{4.4}) has
the dimension of length.

It is convenient to go from Minowsky coordinates (\ref{1.2}) to retarded time coordinates
\begin{equation}\label{4.5}
	ds^2= -dU^2-2dUdr+r^2d\Omega^2~~,
\end{equation}
where $U=t-r$ and $r=\sqrt{x^2+y^2+z^2}$. We denote coordinates $\theta, \varphi$ on the unit sphere by $x^A$, $d\Omega^2=\gamma_{AB}dx^Adx^B=\sin^2\theta d\varphi^2+d\theta^2$.
Our results will be expressed in terms of a unit vector $\vec{n}$ with components $n_x=x/r, n_y=y/r, n_z=z/r$.  If $\vec{l}$ is a unit
vector along the velocity of the string, $\vec{p}$ is another unit vector along the string axis ($l^i=\delta^i_x$, $p^i=\delta^i_z$), then
$n_x=(\vec{n}\cdot \vec{l})$, $n_z=(\vec{n}\cdot \vec{p})$.
We use coordinates (\ref{4.5}) above the string
horizon, $u>0$. In this region $U>-r(1-n_x)$. The region includes the domain $U>0$.

As has been shown in \cite{Fursaev:2022ayo} a null string moving near a point-like charge  disturbs its field
and creates an outgoing EM-pulse traveling
to the future null infinity $\mathcal{I}^{+}$, $r \to \infty$. The trajectory of the pulse is the light cone shown on Fig.~\ref{Cone}.

The energy of the EM field inside the sphere of the radius $R$ with the center at the point $x^i=0$ is
\begin{equation}\label{4.6}
	E(R,t)=\int_{r<R} d^3x~ T_{00}~~,
\end{equation}
where $T^\mu_\nu$ is the stress-energy tensor of the EM field,
\begin{equation}\label{4.7}
	T^\mu_\nu=-F^{\mu\alpha}F_{\nu\alpha}+\delta^\mu_\nu \frac 14 F^{\alpha\beta}F_{\alpha\beta}~~.
\end{equation}
The  energy density $T_{00}$, is measured in the frame of reference where the charge is at rest.
The conservation law implies that
\begin{equation}\label{4.8}
	\partial_t E(R,t)=R^2\int d\Omega~ T^r_U~ ~~.
\end{equation}
We consider the energy flow at $r\gg a$, $r\gg U$. As is shown in the next Section, in the given regime one has the following asymptotics,
which hold for all angles $\Omega$ if $U>0$:
\begin{equation}\label{4.9a}
	A_{S,\mu}(r,U,\Omega)\simeq {\frac{1}{r}}(a_\mu (U,\Omega) + b_\mu (\Omega)\ln r/\varrho)+O(r^{-2}\ln r)~~,~~\mu=U,r~~,
\end{equation}
\begin{equation}\label{4.9c}
	A_{S,B}(r,U,\Omega)\simeq a_B (U,\Omega) + b_B (\Omega)\ln r/\varrho+O(r^{-1}\ln r)~~,
\end{equation}
where $A_{S,B}$ are components in the tangent space to $S^2$, $\varrho$ is a  dimensional parameter related to the approximation.
This yields:
\begin{equation}\label{4.10}
	T^r_U=F_{U\mu}F^{r\mu}\simeq \frac{1}{r^2}\gamma^{AB} \dot{a}_A \dot{a}_B ~~,
\end{equation}
where $\dot{a}_A\equiv\partial_Ua_A$, and guarantees a finite energy flux (\ref{4.8}) at large $r$, for $U>0$,
\begin{equation}\label{4.11}
\lim_{R\to \infty}	\partial_t E(R,t)=\int d\Omega ~\gamma^{AB}\dot{a}_A \dot{a}_B=\int d\Omega \left(\dot{\vec{a}}^2 - (\dot{\vec{a}}\cdot
	\vec{n})^2\right)~~.
\end{equation}
The r.h.s. of (\ref{4.11}) is given for components in Minkowsky coordinates, $n^i=x^i/r$.

The calculation of (\ref{4.11})
is our main goal. In next sections we derive the leading terms in (\ref{4.9a}), (\ref{4.9c})  in an analytic form for sources of electric and magnetic types.

\subsection{Master problem}\label{aux}

As we will see, solutions to vector problem (\ref{1.25}) can be generated by a solution of the following scalar problem:
\begin{equation}\label{5.25}
	\Box \Phi_\omega(u, {\bf x})=0~~,~~\Phi_\omega(0, {\bf x})=\theta(-y)~ f_\omega({\bf x})~~,
\end{equation}
\begin{equation}\label{5.25b}
f_\omega({\bf x})\equiv f({\bar {\bf x}})~~,~~f({\bf x}) \equiv  \phi(v/2,y,z)~~.
\end{equation}
Here ${\bar {\bf x}}$ is defined in (\ref{1.14}), and $\phi(x,y,z)$ in (\ref{coul}). Before we proceed it is helpful
to discuss properties of $\Phi_\omega(x)$ in some detail. The key fact is
the integral representation found in  \cite{Fursaev:2022ayo}:
\begin{equation}\label{5.33}
\Phi_\omega(x)=-C \int_{S^2}d\Omega'~\Re\left(
\frac{\tilde{\Phi}_\omega (\Omega')}{x^\mu\,m_\mu (\Omega') +ia\varepsilon(\Omega')}\right)
	~~,~~\tilde{\Phi}_\omega (\Omega')\equiv\frac{\cos\varphi'}{g(\Omega',\omega)}~~.
\end{equation}
where $C=1 /(8\pi^3)$. The integration goes over a unit sphere $S^2$, with coordinates $\Omega'=(\theta',\varphi')$,
$d\Omega=\sin\theta' d\theta d\varphi'$. Other notations are:
\begin{equation}\label{a.1}
	m_u=1-\sin^2\theta'\cos^2\varphi'~,~m_v=\sin^2\theta'\cos^2\varphi'~,
	~m_y=\sin 2\theta' \cos\varphi'~~,~~
	m_z=\sin^2\theta'\sin 2\varphi'~~,
\end{equation}
\begin{equation}\label{a.4}
g(\Omega',\omega)=e^{i\theta'}+\omega \sin\theta' \cos\varphi'~~,~~\varepsilon(\Omega')=2\sin^2\theta'\cos\varphi'~~.
\end{equation}
One can check that vector field $m_\mu$ is null, $m^2=0$, which guarantees that $\Box \Phi_\omega=0$.

Let $\Phi(x)$ be a solution to (\ref{5.25}), (\ref{5.25b}) for $\omega=0$. The initial data for $\Phi_\omega$
are obtained by the Carroll transformation of initial data for $\Phi$. Let us show that this transformation
generates the null rotation of the solution, that is
\begin{equation}\label{5.51}
\Phi_\omega(x)=\Phi(\bar{x})~~,~~x=M(\omega)\bar{x}~~.
\end{equation}
To this aim we introduce a unit vector on $S^2$
\begin{equation}\label{5.52a}
l_x=\sin\theta'\cos\varphi'~,~l_y=\cos \theta'
~,~l_z=\sin\theta'\sin\varphi'~,
\end{equation}
and express quantities under the integral in (\ref{5.33}) as
\begin{equation}\label{5.52}
m_0=1~~,~~m_x=2l_x^2-1~~,~~m_y=2l_x l_y~~,~~m_z=2l_x l_z~~,~~\varepsilon=2l_x\sqrt{1-l_y^2}~~,
\end{equation}
where $m_0=m_v+m_u$, $m_x=m_v-m_u$.
According to (\ref{5.33})
\begin{equation}\label{5.53a}
\Phi(\bar{x})=\Phi_{\omega=0}(\bar{x})=-C \int_{S^2}d\Omega'~\Re\left(
\frac{\tilde{\Phi} (\Omega')}{x^\mu\,m'_\mu +ia\varepsilon}\right)~~,
\end{equation}
where $m'=M(\omega)m$ and
\begin{equation}\label{5.53b}
\tilde{\Phi} =\tilde{\Phi}_{\omega=0}=\frac{l_x}{\sqrt{1-l^2_y}(l_y+i \sqrt{1-l^2_y})}~~.
\end{equation}
Vector $m'$ can be further transformed to $\tilde{m}=m'/S$
\begin{equation}\label{5.54}
	\tilde{m}_0=1~~,~~\tilde{m}_x=2\tilde{l}_x^2-1~~,~~\tilde{m}_y=2\tilde{l}_x \tilde{l}_y~~,~~
\tilde{m}_z=2\tilde{l}_x \tilde{l}_z~~,
\end{equation}
\begin{equation}\label{5.55}
	\tilde{l}_x=S^{-1/2}l_x~~,~~\tilde{l}_y=S^{-1/2}(l_y-\omega l_x)~~,~~\tilde{l}_z=S^{-1/2}l_z~~,~~S=1-2\omega l_xl_y+
	\omega^2 l_x^2~~.
\end{equation}
This yields
\begin{equation}\label{5.56}
\frac{\tilde{\Phi}}{(x')^\mu\,m'_\mu +ia\varepsilon}=S^{-3/2}\frac{\tilde{\Phi}_\omega}{(x')^\mu\,\tilde{m}_\mu +ia\tilde{\varepsilon}}~~,
\end{equation}
where $\tilde{\Phi}_\omega,\tilde{\varepsilon}$ are determined in terms of $\tilde{l}_i$. Then
(\ref{5.51}) easily follows from (\ref{5.53a}), (\ref{5.53b}) if one changes the integration measure.

\subsection{Radiation from an electric charge}\label{elrad}

For an electric source perturbation $A_{S,b}$ is a solution to (\ref{1.25}) with
\begin{equation}\label{5.1}
\bar{a}_b({\bf x})=\frac12 f({\bf x}) \, \delta_b^v~~,
\end{equation}
where  $f({\bf x})$ is defined in (\ref{5.25b}). Equation (\ref{5.1}) is the consequence of (\ref{4.2}), (\ref{coul}).
As is easy to see, $A_{S,b}$ is generated by the solution to scalar problem (\ref{5.25}) in the following way:
\begin{equation}\label{5.50}
A_{S,v}(x)=\frac e2 (\Phi_\omega(x)-\Phi(x))~~,~~A_{S,y}(x)=-e\omega \Phi_\omega(x)~~,~~A_{S,z}(x)=0~~.
\end{equation}
The rest component, $A_{S,u}$, is determined by the gauge condition $\partial A_{S}=0$.
The solution at $u>0$ can be represented as \cite{Fursaev:2022ayo}
\begin{equation}\label{5.2}
A_{S,\mu} (x)=-eC \int_{S^2} d\Omega'~\Re\left[\frac{\beta_\mu (\Omega')}{ x^\nu m_\nu (\Omega')+ia\varepsilon (\Omega')}\right]
~~.
\end{equation}
The new notations used in (\ref{5.2}) are:
\begin{equation}\label{a.2}
\beta_v=-\frac 12 \cos\varphi'~\left(g^{-1}(\Omega',\omega)-g^{-1}(\Omega',0)\right)
	~,~~\beta_y=\cos\varphi'~\omega  g^{-1}(\Omega',\omega)~~,~~\beta_z=0~,
\end{equation}
\begin{equation}\label{a.3}
\beta_u=\frac{m_y\beta_y-2m_u \beta_v}{2m_v}~~,
\end{equation}
with $g(\Omega',\omega)$, $\varepsilon (\Omega')$ defined in (\ref{a.4}).  The gauge condition, which results in
(\ref{a.3}) allows residual gauge transformations. So solution to $\beta_u$ is not unique. This arbitrariness, however, does not affect physical observables.

In coordinates $U,r,x^A$, see (\ref{4.5}), the denominator in the integral in (\ref{5.2}) can be written as
$$
x^\nu m_\nu +ia\varepsilon =U+r((\vec{m}\cdot \vec{n})+1)+ia\varepsilon~~~.
$$
Since we are interested in a large $r$ asymptotic,
the integration in (\ref{5.2}) can be decomposed into two parts: the integration  over a domain, where the factor $(\vec{m}\cdot \vec{n})+1$ is small,
that is, $\vec{m}$ is almost $- \vec{n}$,  and the integration over the rest part of $S^2$. If we introduce a dimensionless parameter
$\Lambda$ such that
\begin{equation}\label{5.6}
\frac{\sqrt{U^2+a^2}}{ r} \ll \Lambda^2 \ll 1~~,
\end{equation}
the first region can be defined as $(\vec{m}\cdot \vec{n})+1\leq \Lambda^2$, and the second  as
$(\vec{m}\cdot \vec{n})+1>\Lambda^2$. Contributions from these regions to $A_{S}$ will be denoted as $A_1$ and
$A_2$, respectively,
\begin{equation}\label{5.10a}
A_{S}= A_{1}+A_{2}~~.
\end{equation}
After some algebra one gets the following estimates at large $r$ (for components in coordinates (\ref{1.2})) :
\begin{equation}\label{5.7}
A_{1,\mu}(r,U,\Omega)\simeq  \frac{eN(\Omega)}{r}\Re~\left[\bar{\beta}_\mu\ln
\left(\frac{U+ia\bar{\varepsilon}+r\Lambda^2}{ U+ia\bar{\varepsilon}}\right)\right]~~,
\end{equation}
\begin{equation}\label{5.9}
A_{2,\mu}(r,U,\Omega)\simeq  \frac{b_{2,\mu}(\Omega,\Lambda)}{r}~~,
\end{equation}
\begin{equation}\label{5.8}
\bar{\beta}_\mu=\beta_\mu \mid_{\vec{m}=-\vec{n}}~~,~~
\bar{\varepsilon}=\varepsilon\mid_{\vec{m}=-\vec{n}}~~,
\end{equation}
\begin{equation}\label{5.8b}
N(\Omega)=\frac{1}{4\pi^2}\sqrt{\frac{2}{1-n_x}}~~,
\end{equation}
\begin{equation}\label{5.9b}
b_{2,\mu}(\Omega,\Lambda)=-eC
\int_{S^2_\Lambda} d\Omega'~\cos\varphi~\Re\left[\frac{\beta_\mu (\Omega')}{ (\vec{m}(\Omega')\cdot \vec{n})+1}\right]~~,
\end{equation}
where $S^2_\Lambda$ is a part of $S^2$
with the restriction $(\vec{m}\cdot \vec{n})+1>\Lambda^2$.  Arguments which lead to (\ref{5.7}) can be found in Appendix \ref{App1}.
Calculation of (\ref{5.9}) is trivial.

As a result of (\ref{5.7}),  (\ref{5.9}) the solution at large $r$ is a sum of a static, $U$ independent part,
and a dynamical part. The dynamical part presents only in $A_1$ and it is the only part
which contributes to the flux (\ref{4.11}). One can show that formula (\ref{5.7}) is in a good agreement with numerical simulations for $n_x \neq 1$, so it provides a remarkable analytic tool to compute different characteristics of the EM field at future null infinity.
We come to the following expression which leads to Eq. (\ref{i.1}) announced in Sec. \ref{intr}:
\begin{equation}\label{5.10}
A_{S,\mu}(r,U,\Omega)\simeq \frac{a_{\mu}(U,\Omega)}{ r} +\frac{b_{\mu}(\Omega)}{ r} \ln{\varrho / r}~~,
\end{equation}
\begin{equation}\label{5.12}
a_{\mu}(U,\Omega) =-eN(\Omega)~ \Re~\left[\bar{\beta}_\mu\ln
\left(\frac{U+ia\bar{\varepsilon}}{ a}\right) \right]+b_{2,\mu}(\Omega,\Lambda)~~,
\end{equation}
\begin{equation}\label{5.11}
b_{\mu}(\Omega,\Lambda) =eN(\Omega)~\Re~ \bar{\beta}_\mu (\Omega)~~,
\end{equation}
where $\varrho=a/\Lambda^2$.

The power of the radiation
is given by (\ref{4.11}).  The dynamical part is determined by $\bar{\beta}_\mu$, with the condition
$\bar{\beta}^\mu \bar{m}_\mu =0$. It implies that $\dot{a}_0=-(\vec{\dot{a}}\cdot \vec{n})$ and
\begin{equation}\label{5.13}
\dot{E}=\int d\Omega ~\gamma^{AB}\dot{a}_A \dot{a}_B=\int d\Omega \left(\dot{a}_\mu\eta^{\mu\nu}\dot{a}_\nu\right)\equiv
\int d\Omega ~f_E(U,\Omega)~~,
\end{equation}
where $\eta_{\mu\nu}$ is the flat metric.
If we express $\dot{a}_v$ through the rest components by using the gauge condition $\dot{a}_\mu m^\mu=0$, where $m^u=-(1-n_x)$,
$m^v=-(1+n_x)$, $m^y=-n_y$,
the flux density (intensity of the radiation) takes the form:
\begin{eqnarray}\label{5.16.0}
f_E(U,\Omega)&=&
\dot{a}_\mu\eta^{\mu\nu}\dot{a}_\nu=
\frac{4 n_y \dot{a}_y\dot{a}_v+4(1+n_x)\dot{a}_v^2+(1-n_x)\dot{a}_y^2}{ 1-n_x}~~.
\end{eqnarray}
Unit vector $\vec{n}$ is introduced after Eq. (\ref{4.5}).
The functions $\dot a$ are expressed in terms of real and imaginary parts
of $\bar{\beta}_\mu$,
\begin{eqnarray}\label{5.14a}
	\dot a_{\mu}=-
	\frac{e N(\Omega)}{U^2+a^2 \bar \varepsilon^2} \left[U \,\Re\bar\beta_{\mu}+a\bar\varepsilon\, \Im\bar \beta_{\mu}\right],
\end{eqnarray}
see (\ref{5.12}). We are interested in the limit when the string energy is small, $\omega\ll 1$. With the help of  (\ref{a.10})-(\ref{a.14}) one gets  in the linear approximation in $\omega$
 \begin{eqnarray}
&& \Re \bar\beta_{y}\simeq -\frac{\omega n_y}{\sqrt{2}}\frac{\sqrt{1-n_x}}{\bar\varepsilon},\quad  \Re \bar\beta_{v}\simeq\frac{\omega}{2\sqrt{2}}\frac{\sqrt{1-n_x}}{\bar\varepsilon}(n_y^2-(1-n_x)), \nonumber\\
&&  \Im \bar\beta_{y}\simeq-\frac{\omega}{\sqrt{2}}\sqrt{1-n_x},\quad  \Im \bar\beta_{v}\simeq\frac{\omega n_y}{2\sqrt{2}} \sqrt{1-n_x}~~.
\label{5.15}
\end{eqnarray}
Substitution of  (\ref{5.14a}), (\ref{5.15}) to (\ref{5.16.0}) yields
\begin{eqnarray}\label{5.16.a}
f_E(U,\Omega)&\simeq&\frac{e^2 \omega^2}{(2\pi)^4 a^2}\frac{1}{[U^2/a^2+\bar\varepsilon^2]^2} \\
&&\times\left\{\frac{1-n_x^2-n_y^2(1-\bar\varepsilon^2)}{\bar\varepsilon^2}\,\frac{U^2}{a^2}+2n_y(1-n_x-n_y^2) \frac{U}{a}+(1-n_y^2)\bar\varepsilon^2 \right\}, \nonumber
\end{eqnarray}
where $\bar\varepsilon^2=2(1-n_x)-n_y^2$, $n_x=\cos\theta$, $n_y=\sin\theta\sin\varphi$. We consider (\ref{5.16.a}) at $U>0$.

\begin{figure}
	\begin{center}
		\includegraphics[width=11cm]{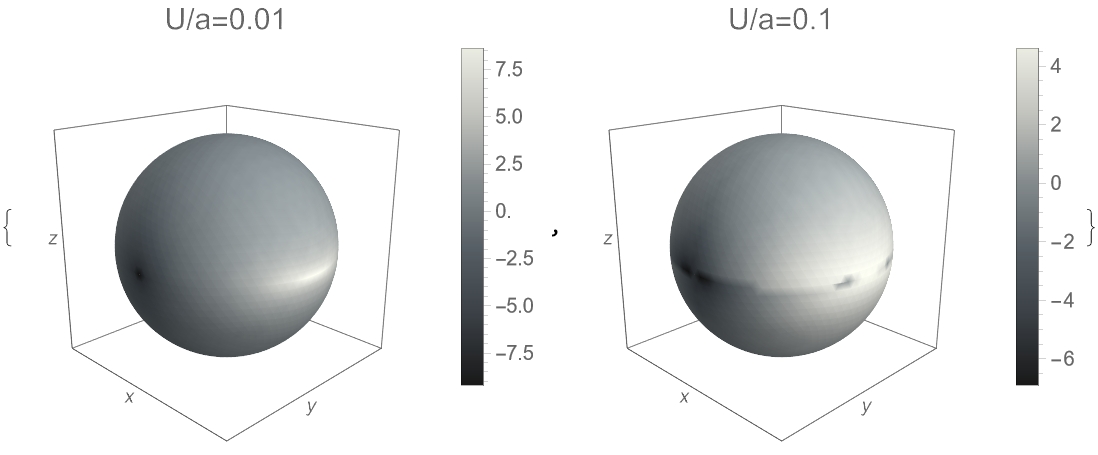}
	\end{center}
	\caption{The flux density (divided by  $e^2\omega^2/((2\pi)^4 a^2)$) from an electric source. The flux
is given in the logarithmic scale in $4\pi$ geometry for $U=0.01, 0.1$ and impact parameter $a=1$. The radiation has a form of EM burst which is directed mostly toward the velocity of the string and rapidly decays with increasing of $U$.}
\label{fig2}
\end{figure}

For $U/a\gg 1$ the flux density vanishes as $(U/a)^{-2}$. It follows from (\ref{5.12}) and numerical simulations in \cite{Fursaev:2022ayo} that the peak power is near $U=0$. Such behavior is demonstrated on Fig. \ref{fig2} which shows an angular distribution of the
flux (intensity of the flux) at different moments of $U$. For small $U$ the energy flux is focused in the direction of string motion.
For $n_x\to\pm 1$  and finite $U$,
\begin{equation}
f_E(U,\Omega)_{n_x\to1}=\frac{e^2}{(2\pi)^4}\frac{\omega^2}{U^2}, \quad f_E(U,\Omega)_{n_x\to-1}=\frac{e^2}{16\pi^4}\frac{4\omega^2}{(4+U^2)^2}~~~.
\end{equation}
At  $U=0$, expression  (\ref{5.16.a}) takes the simple form
\begin{equation}\label{5.16b}
f_E(0,\Omega)=\frac{e^2}{(2\pi)^4} \frac{\omega^2}{ a^2}
   \frac{(1-n_y^2)}{2(1-n_x)-n_y^2}~~.
\end{equation}
It has a pole-like singularity at $n_x=1$, where our approximation is not applicable. Note that at $U=0$ this point lies exactly on the string world-sheet $\cal S$.

\begin{figure}[t]
	\begin{center}
		\includegraphics[height=4cm]{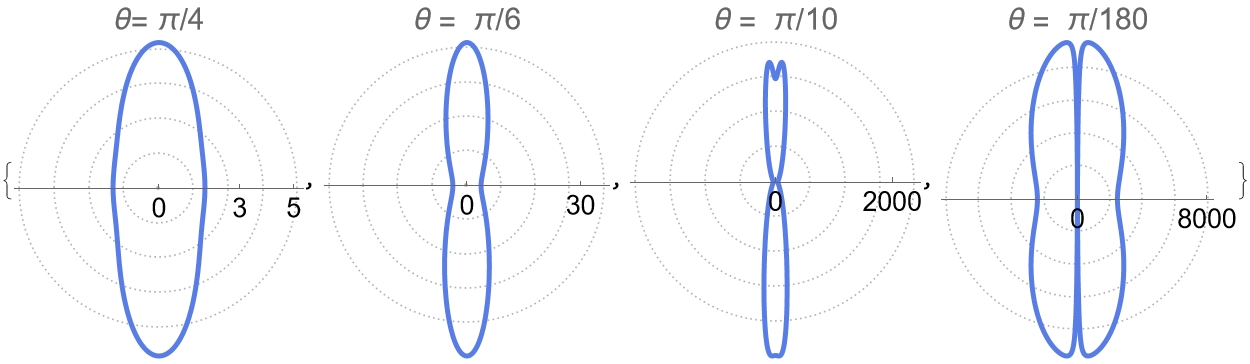}
	\end{center}
	\caption{Polar radiation plots (far-field patterns) in the planes $\theta=\pi/4, \pi/6, \pi/10,\pi/180$.
		The flux density is depicted for  $U/a=0.01$ when  $e^2\omega^2/((2\pi)^4 a^2)=1$. Angles $\theta$ and $\varphi$ are defined by
(\ref{a.5}). Angle $\varphi$ changes clock-wise. The intensity of the flux grows in the direction of the string velocity, as $\theta\to 0$. The flux is orthogonal to the axis of the string, which corresponds to $\varphi=0,\pi$.}
		\label{fig3}
\end{figure}

To better illustrate the intensity of the flux we present on Fig. \ref{fig3} polar radiation plots   which are common in antenna physics. 
 They refer to the directional (angular) dependence of the flux density of the EM-waves coming from  an electric source. The flux density \eqref{5.16.a} depends on two angles and retarded Bondi coordinate $U$.  We fix the distance from the source, $U=0.01$, and one  of the  angels, $\theta$, and draw the flux density in polar coordinates. Then the polar radiation graphs in the  Fig. \ref{fig3} show the dependence of the flux density on the angle $\varphi$ in different planes $\theta=\mbox{const}$,  and $e^2\omega^2/((2\pi)^4 a^2)=1$.  We return to an estimate of the flux from a null  cosmic string and
a gas of charged particles  in Sec. \ref{exp}.

\subsection{Radiation from a magnetic-dipole source}\label{marad}

The case of point-like magnetic-dipole sources has not been studied in \cite{Fursaev:2022ayo}. Therefore, we consider this case in more detail
by taking into account results of \cite{Fursaev:2022ayo}. It follows from (\ref{4.4})  that incoming data
for problem (\ref{1.25}) for a magnetic-dipole source are
\begin{equation}\label{5.19}
\bar{a}_b({\bf x})={\mathcal D}_b f({\bf x})~~,~~b=v,y,z~~,
\end{equation}
where  $f({\bf x})$ is defined in (\ref{5.25b}), and
\begin{equation}\label{5.20}
2{\mathcal D}_v=(M_y \partial_z-M_z \partial_y), \quad {\mathcal D}_y= -M_x \partial_z+2 M_z \partial_{v}, \quad {\mathcal D}_z=M_x \partial_y-{\color{black}2}M_y \partial_{ v}~~.
\end{equation}
If we introduce the operators:
\begin{equation}\label{5.21} 
{\color{black}{\mathcal D}^{\omega}_b={\mathcal D}_b +\omega(-M_z \delta_b^v+2 M_x \delta^z_b)\partial_v } ~~,
\end{equation}
then, according to (\ref{1.20}), the initial data for perturbations induced by the string can be written as
\begin{eqnarray}\label{5.22}
{\color{black} a_{b}({\bf x})}&=& {\color{black}\theta(-y) \left({\mathcal D}^\omega_{b}f_\omega ({\bf x})-
2\omega \delta_b^y {\mathcal D}^\omega_{v}f_\omega ({\bf x})- {\mathcal D}_{b}f({\bf x})\right)},
\end{eqnarray}
Here $f_\omega({\bf x})$ is introduced in (\ref{5.25b}) and operators ${\mathcal D}^\omega_{b}$ are obtained from ${\mathcal D}_{b}$ by replacing $\partial_y$ to
$\partial_y+2\omega \partial_v$ since
$(\partial_y+2\omega \partial_v)f_\omega ({\bf x})=\partial_y f({\bf x})_{{\bf x}=\bar{{\bf x}}}$.

Let $\Psi^\omega_b(u, {\bf x})$ be a solution at $u>0$ of the
following problem:
\begin{equation}\label{5.23}
\Box \Psi^\omega_b(u, {\bf x})=0~~,~~\Psi^\omega_b(0, {\bf x})=\theta(-y)~{\mathcal D}^\omega_ b f_\omega
({\bf x})~~,~~b=v,y,z~~.
\end{equation}
Then the solution to (\ref{1.25}) for the magnetic-dipole source has the components:
\begin{eqnarray}\label{5.24}
{\color{black}A_{S,b}(x)}&=&{\color{black}\Psi^\omega_b(x)-\Psi_{b}(x) -2\omega  \delta_b^y \Psi^\omega_v(x)~~.  }
\end{eqnarray}
where $\Psi_{b}(x)=\Psi^{\omega=0}_b(x)$. To proceed we consider another auxiliary  problem
\begin{equation}\label{5.26}
\Box \chi(u, {\bf x})=0~~,~~\chi(0, {\bf x})=\delta(y)~ f({\bf x})~~.
\end{equation}
Given solutions to (\ref{5.25}), (\ref{5.26}) one can easily construct the solution
\begin{equation}\label{5.27}
\tilde{\Psi}^\omega_b(x)=\partial_b \Phi_\omega+\delta_b^y\chi(x)~~,
\end{equation}
to the problem:
\begin{equation}\label{5.28}
\Box \tilde{\Psi}^\omega_b(u, {\bf x})=0~~,~~\tilde{\Psi}^\omega_b(0, {\bf x})=\theta(-y)~ \partial_b f_\omega ({\bf x})~~,
\end{equation}
and represent $\Psi^\omega_b$ as
\begin{equation}\label{5.29}
\Psi^\omega_b(x)={\mathcal D}^\omega_ b\Phi_\omega(x)+(M_x\delta^z_b-{\color{black}2}M_z\delta^v_b)\chi(x)~~.
\end{equation}
Now the final result for components can be written with the  help
of (\ref{5.24}), (\ref{5.29})  in the form:
\begin{eqnarray}\label{5.30a}
{\color{black}A_{S,a}(x)}&=&{\color{black}{\mathcal D}^\omega_ a\Phi_\omega(x)-{\mathcal D}_ a\Phi(x) -2\omega \bigl({\mathcal D}^\omega_ v \Phi_\omega(x)-2 M_z \chi(x)\bigr)\delta^a_y},
\end{eqnarray}
where $\Phi(x)=\Phi_{\omega=0}(x)$.  Equation (\ref{5.30a}) is our starting point for computations of the flux.

By taking into account (\ref{5.30a}) and the gauge conditions
perturbation of the vector-potential of the magnetic source caused by the null string can written as:
\begin{equation}\label{5.39}
A_{S,\nu}(x) =A_{1,\nu}(x) +A_{2,\nu}(x)~~~.
\end{equation}
By its structure $A_{1,\nu}(x)$ is analogous to potential (\ref{5.2}) for an electric source,
\begin{equation}\label{5.40}
A_{1,\nu}(x) =C \int_{S^2}d\Omega'~\Re\left\{
\frac{\alpha_\nu (\Omega')}{ (x^\mu\,m_\mu +ia\varepsilon)^2}\right\}
~~,
\end{equation}
\begin{eqnarray}
\label{5.41}
{\color{black}\alpha_a}(\Omega)&=&{\color{black} m_{\mu}(({\mathcal D}_a^{\omega}x^{\mu})-2\omega \delta_a^y ({\mathcal D}_v^{\omega} x^{\mu})) \tilde{\Phi}_\omega - m_{\mu}({\mathcal D}_a x^{\mu}) \tilde{\Phi} }, \quad a=v,y,z \\
\label{5.42}
\alpha_u(\Omega)&=&\frac{m_y \alpha_y+ m_z \alpha_z-2 m_u \alpha_v}{2m_v}~,
\end{eqnarray}
where $\tilde{\Phi}_\omega (\Omega)={\cos\varphi  g^{-1}(\Omega,\omega)}$, see (\ref{5.33}).  The rest part of the solution
presents only if $M_z\neq 0$ and  can be written as
\begin{equation}\label{5.43}
A_{2,\nu}(x) ={\color{black}2} \omega M_z(2\delta_\nu^y \partial_v+\delta_\nu^u \partial_y)\bar{\chi}
~~.
\end{equation}
Here $\partial_v\bar{\chi}=\chi$. Properties of function $\chi$ are studied in detail in  Appendix \ref{App2}.
The $u$-component of $A_\mu$ in (\ref{5.42}), (\ref{5.43}) is determined by the gauge condition $\partial A=0$. As has been pointed out
the gauge condition fixes the solution up to residual gauge transformations, which do not change physical observables, like the energy flux.

With the help of (\ref{5.33}) one finds the asymptotic form of (\ref{5.40}) for $n_x\neq 1$
\begin{equation}\label{5.44}
A_{1,\mu} (x)\simeq \frac{N(\Omega)}{r} ~
\Re\left(\frac{\bar{\alpha}_\mu}{U+ia\bar{\varepsilon}} \right)~~,
\end{equation}
\begin{equation}\label{5.45}
\bar{\alpha}_\mu=\alpha_\mu \mid_{\vec{m}=-\vec{n}}~~.
\end{equation}
Function $N$ is defined in (\ref{5.8b}), and, as earlier, $\bar\varepsilon^2=2(1-n_x)-n_y^2$. Components of $\bar{\alpha}_\mu$ follow from  (\ref{5.41})-(\ref{5.42}) where one
has to replace $\tilde{\Phi}_\omega$ to
\begin{equation}\label{5.46}
\bar{\Phi}_\omega =\frac{\sqrt{2(1-n_x)^3}}{\bar{\varepsilon}(-n_y+\omega(1-n_x)+i\bar{\varepsilon}) }~~,
\end{equation}
Derivation of  (\ref{5.44}) can be performed along the lines of electric case elaborated in Appendix \ref{App1}.
Asymptotic of $\chi$ at large $r$, which is derived in Appendix \ref{App2}, is
\begin{equation}\label{5.47}
\bar{\chi}(x)=\frac{1}{ 2\pi^2 r}\left[\frac{1 }{4\bar{\varepsilon}}\ln \left(\frac{\sqrt{U^2 +a^2 \bar{\varepsilon}^2}}{ 2Lr} \right)-X(\Omega)\right].
\end{equation}

Like in (\ref{5.13}), we calculate the density of the energy flux, or the intensity, $f_M(U,\Omega)$,
\begin{equation}\label{5.13.m}
\dot{E}\equiv
\int d\Omega ~f_M(U,\Omega)~~,
\end{equation}
\begin{equation}\label{5.49m}
f_M(U,\Omega)=\frac{4(n_y \dot a_y \dot a_v+n_z \dot a_z \dot a_v+(1+n_x)\dot a_v^2)+(1-n_x)(\dot a_y^2+\dot a_z^2)}{(1-n_x)}~,
\end{equation}
where $\dot a_{\mu}$ is defined by the asymptotic $\dot A_{\mu} (x)\simeq  \dot a_{\mu}/r $ at large $r$. The difference
of this expression with respect to the case of the electric source is in non-vanishing component $\dot a_z$.
After some algebra  one arrives at the following result valid at small $\omega$:

			\begin{eqnarray} \label{5.48a}
		\dot a_{b}&\simeq &-\frac{\omega}{8\pi^2 a^2\bar\varepsilon}\, \frac{\sigma_i\zeta_{b}^i
		}{((U/a)^2+\bar\varepsilon^2)^2}~~,\quad i=1,2,3,\\	
	\sigma_1&=&	(n_y^2-\bar\varepsilon^2)((U/a)^2-\bar\varepsilon^2)+4 \bar\varepsilon^2 n_y (U/a)\\
	\sigma_2&=& 2 \bar\varepsilon( n_y ((U/a)^2-\bar\varepsilon^2)-(n_y^2-\bar\varepsilon^2)(U/a))\\
	\sigma_3&=& (U/a)^2-\bar\varepsilon^2	
	\end{eqnarray}	
{\color{black}
		\begin{eqnarray*} \label{5.48}
	&&\zeta_{v}^1=\frac{1}{2} M_y n_z, \quad \zeta_{z}^1=M_y(1-n_x), \\
&&\zeta_{y}^1=-(M_x n_z+M_z(1-n_x))-\frac{n_y}{1-n_x}(M_y n_z-M_z n_y), \\
	&&	\zeta_{v}^2=\frac{\bar\varepsilon}{2}M_z, \quad 	\zeta_{z}^2=-\bar\varepsilon M_x,\quad
	\zeta_{y}^2=-\frac{\bar\varepsilon}{1-n_x}(M_y n_z-M_z n_y);\\
&&	\zeta_{b}^3=2 M_z(1-n_x)\delta_b^y;
	\end{eqnarray*}
}


Here the presence of $\zeta_{b}^3$ in (\ref{5.48}) is related to the part of the solution defined in (\ref{5.43}).
The intensity of the radiation $f_M(U,\Omega)$ can be obtained with the help of (\ref{5.49m}), (\ref{5.48a}).

\begin{figure}
	\begin{center}
		\includegraphics[width=14cm]{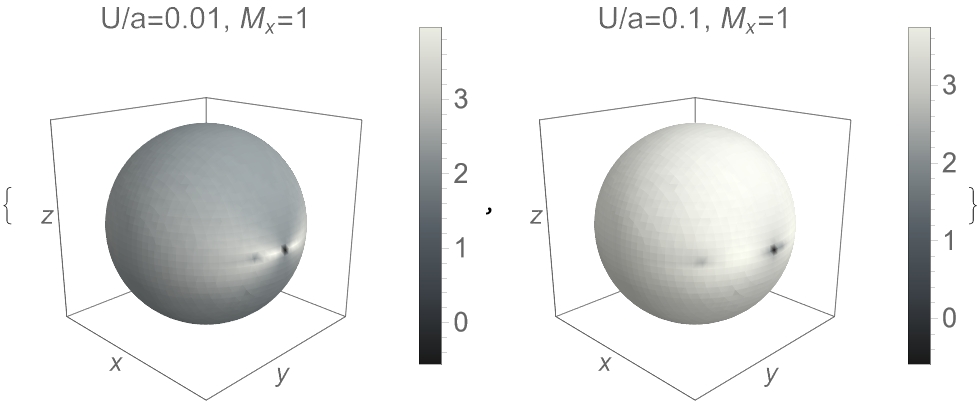}
		\includegraphics[width=14cm]{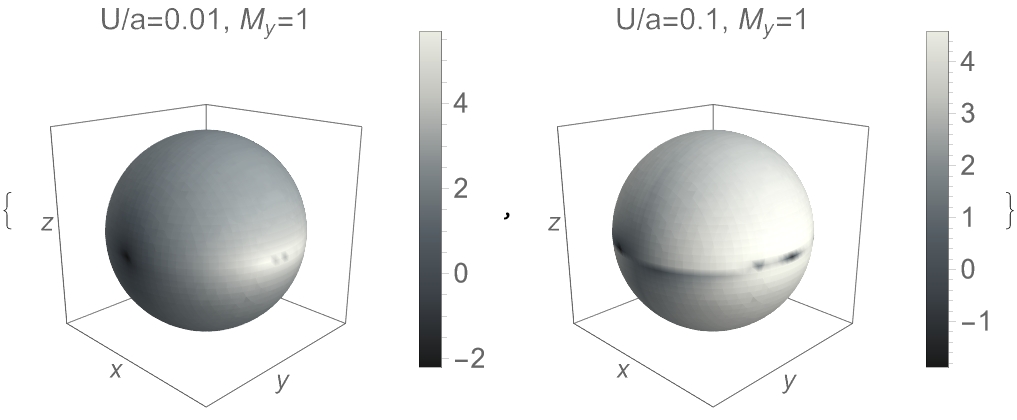}
		\includegraphics[width=14cm]{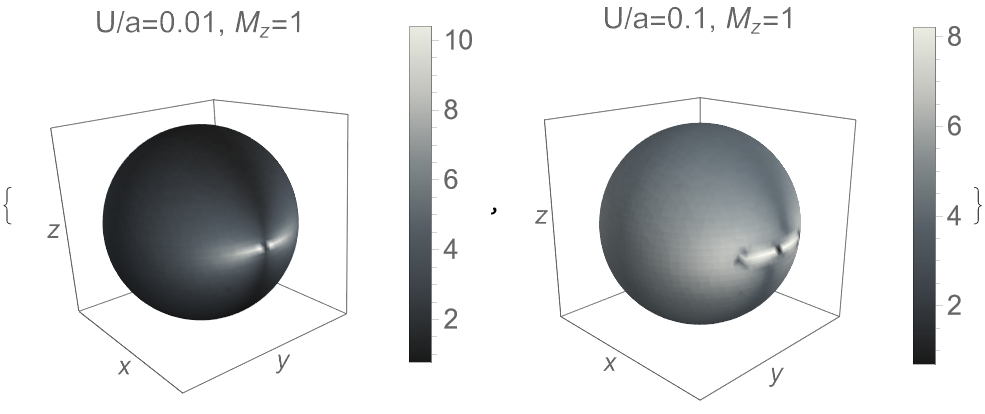}
	\end{center}
	\caption{shows the flux density (divided by $\omega^2/(64\pi^4 a^4)$) from a magnetic-dipole source for  $U/a=0.1$ and $U/a=0.01$, $|M|=1$, in logarithmic scale. Three different directions of the magnetic moment are chosen: toward the string velocity, $M_x=1$, orthogonally to the string
		and its velocity, $M_y=1$, and along the string, $M_z=1$.}
		\label{flow_M}
\end{figure}

Similarly to the case of the electric source $f_M(U,\Omega)$  is distributed predominantly in the direction of the string motion, however now
it depends on the direction of the magnetic moment $\vec M$ with respect to the string.
Figure \ref{flow_M} shows the angular distribution of the flux density over a unit sphere for different directions of $\vec M$ in logarithmic scale.
The highest flux density is achieved when the magnetic moment is aligned with the sting, $M=M_z$;
the weakest effect is observed when the  magnetic moment is orthogonal both to the string and to its velocity, $M=M_y$.

For illustrative purposes we give concrete expressions of the flux for some special cases at small $\omega$.
At finite $U>0$ one can conclude that {\color{black}in the given approximation  the flux in the direction  $n_x=1$ tends to zero}. 
This direction corresponds to an observer with coordinates $x_o=r,y_o=z_o=0$, that is observer  between the string and the source on trajectory of the string.  It should be noted that  at $n_x=1$ our approximation \eqref{a2.16} is violated.

Another observer on the same trajectory positioned such that the source is between the observer and the string,
with coordinates $x_o=-r,y_o=z_o=0$, sees the flux which depends on the direction of $\vec{M}$ :
\begin{eqnarray}\label{5.49_1}
\vec M=(0,0,M)&&f_{M}(U)={\color{black}\frac{9}{16}} \frac{\omega^2 M^2}{\pi^4 a^4}\frac{(U^2/a^2-4)^2}{(U^2/a^2+4)^4}~~,\\
~\vec M=(0,M,0)&&f_{M}(U)= {\color{black}\frac{1}{4}}\frac{\omega^2 M^2}{ \pi^4 a^4}\frac{(U^2/a^2-4)^2}{(U^2/a^2+4)^4}~~,\\
\vec M=(M,0,0)&&f_{M}(U)= \frac{4\omega^2 M^2}{\pi^4 a^4}\frac{U^2/a^2}{(U^2/a^2+4)^4}~~,~~
\end{eqnarray}
 If $U=0$, and $n_x\neq 1$ one finds for $\vec M=(0,0,M)$,
  {\color{black}
 \begin{eqnarray}\label{5.49_z}
 f_{M}(0,\Omega)&=&
 \frac{\omega^2 M^2}{16 \pi^4 a^4  \bar \varepsilon^6}  \left\{(2-n_x)^2 (1-n_x)^2+ 2(1- n_x)(6+2 n_x-n_x^2)n^2_y \right.
\\
&& \left. +(n_x^2+2 n_x-8)   n_y^4 + n_y^6 \right\};
 \nonumber
 \end{eqnarray}
} 
%
for $\vec M=(0,M,0)$,
 {\color{black}
\begin{eqnarray}\label{5.49_y}
f_{M}(0,\Omega)&=& \frac{\omega^2 M^2}{16 \pi^4 a^4  \varepsilon^2}(1-n_y^2);
\end{eqnarray}
}
%
and for $\vec M=(M,0,0)$,
 {\color{black}
\begin{eqnarray}\label{5.49_x}
f_{M}(0,\Omega)&=& \frac{\omega^2 M^2}{16 \pi^4 a^4 \bar \varepsilon^6} (1 - n_x)^2 (1 - n_x^2 +  (1- 2 n_x) n_y^2 - n_y^4)
\end{eqnarray}
}

At $n_x\to 1$ the flux density diverges since the observer
intersects  trajectory of the string.

\begin{figure}[t]
	\begin{center}
			\includegraphics[height=4cm]{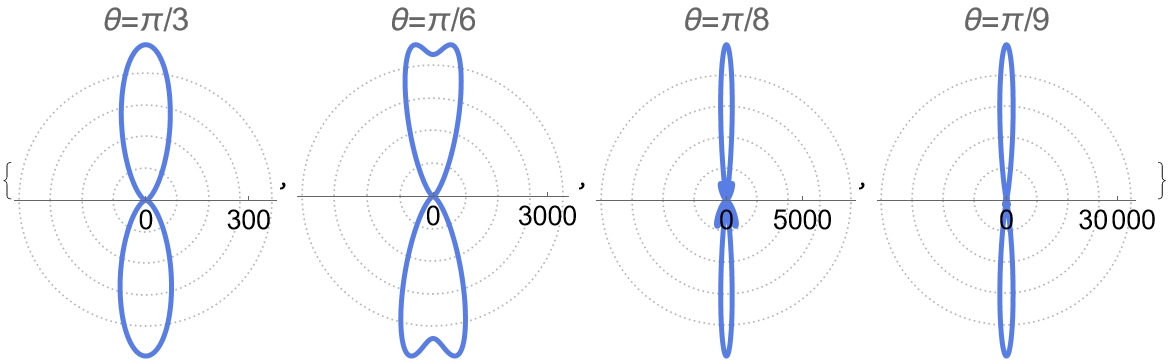}
	\end{center}
	\caption{shows polar radiation plots (far-field patterns)  in the planes $\theta=\pi/3, \pi/6,\pi/8,\pi/9$ for an MD source, for
$U/a=0.1$, when  $e^2\omega^2/((2\pi)^4 a^2)=1$. The magnetic moment is directed along the string. Angles $\theta$ and $\varphi$ are defined as for Fig. \ref{fig3}, $\varphi$ changes clock-wise. Like in case of the electric source the intensity of the flux grows in the direction of the string velocity $\theta=0$. The flux is orthogonal to the axis of the string.}
			\label{polar_M}
\end{figure}

For the case of maximal intensity, when  the magnetic moment is parallel to the string axis, $M=M_z=1$, the polar radiation plot of the flux density  divided by  $\omega^2/(64\pi^4 a^4)$ is shown on  Figure \ref{polar_M} for $U/a=0.1$ . 
 The angle $\phi$ changes from $0$ to $2\pi$ clock-wise. The direction along the string velocity (along the $x$ axes) is $\phi=\pi/2$.
The plots are given for different angular positions of the  observer: $\theta=\pi/3, \pi/6,\pi/8,\pi/9$.  As $\theta$ decreases the plot is stretched along  the $x$ axis. Thus, the radiation is direction toward the motion of the string.  {\color{black} 
Note that at $U/a=0.01$ (the value at which similar graphs are plotted in the case of an electric source), the graphs stretch in the direction of the string velocity by several orders of magnitude.
}
%

\section{Toward experimental observations of EM radiation generated by null strings}\label{exp}
\setcounter{equation}0

To understand
at which energies of null strings the pulses can be potentially observable we consider the  power  of the generated radiation. This power can be considered as the luminosity of the system
which consists of the charge and the null string.   We take $0<U/a \ll 1$ for the peak power. As the preceding analysis shows, the flux densities are maximal at these values.

Let us start with the radiation from electrically charged sources. Formulae
(\ref{5.13}), (\ref{5.16.a})  yield the crude estimate for the luminosity:
\begin{equation}\label{6.1}
\dot{E}\sim \frac{e^2}{(2\pi)^4}~ \frac{\omega^2}{a^2}~~.
\end{equation}
There is an upper limit on the tension of tensile strings $G\mu \leq 10^{-7}$  which follows from CMB constraints
\cite{Planck:2013mgr}-\cite{Dvorkin:2011aj}. Since the effects of null strings on the CMB spectrum are similar to those
of the tensile strings  \cite{Fursaev:2017aap} we use the energy parameter $\omega_0=1\cdot  10^{-7}$ as some reference value to proceed with computations. With the help of this parameter the peak luminosity of the radiation emitted by $N$ charged particles can be written as:
\begin{equation}\label{6.2}
\dot{E}\sim 0.6~\left(\frac{\omega}{\omega_0}\right)^2 \left(\frac{1~\mbox{cm}}{a}\right)^2
~ \frac{N}{ N_A}~\frac{\mbox{erg}}{\mbox{s}}~~,
\end{equation}
Here $N_A$ is the Avogadro number, and we assumed that $e^2/(4\pi)\simeq 1/137$.

Suppose a null string with energy
$\omega=\omega_0$ moves through 1 mole of ionized Hydrogen atoms being under normal conditions. For the mole volume 14 cm$^3$ one can assume $a\simeq 1$ cm and get from (\ref{6.2}) the estimate:
\begin{equation}\label{6.3}
\dot{E}\sim 1~\frac{\mbox{erg}}{ \mbox{s}}=1\cdot 10^{-7} ~W~~.
\end{equation}
Corrections to (\ref{6.2}) related to motion of particles with velocity $v$ is negligible since under normal conditions $v/c \sim 10^{-6}$.
Power (\ref{6.3}) is extremely small. It is just 2 orders of magnitude larger than the power of radiation of a star which goes through  1 square meter of the Earth surface. Thus, radiation from charged sources generated by null cosmic strings may have rather theoretical interest.

Consider now the radiation from magnetic-dipole sources.  Equations
(\ref{5.49m}), (\ref{5.48})  yield the following estimate for the luminosity at small $U$:
\begin{equation}\label{6.4}
\dot{E} \sim \frac{1}{ (2\pi)^4}~ \frac{\omega^2 M^2}{ a^4}~~.
\end{equation}
To proceed we need a physical source with a large magnetic moment $M$ whose field can be approximated by magnetic-dipole potential (\ref{4.4}).  As such sources we consider pulsars, whose electromagnetic fields in a near zone \cite{Beskin:2013UFN} have form (\ref{4.4}).
The near zone of a pulsar is defined by distances from the source $r < c/\Omega$, where $\Omega$ is the pulsar rotation frequency.
Say, if we take  $\Omega\sim 200 \, s^{-1}$,  the near zone is $r< r_0=3.29\cdot 10^2$ km.

A typical value of the surface magnetic field of a pulsar is $B_0\sim 10^{12}~ G$, which corresponds to a magnetic moment $M_0=4.17\cdot 10^{30} \mbox{erg}/G=3.7\cdot 10^{52}$~GeV$^{-1}$, see \cite{Gunn:1969}. One can write (\ref{6.4}) roughly as
\begin{equation}\label{6.4b}
\dot{E} \sim  3.4\cdot 10^{21}~\left(\frac{\omega}{ \omega_0}\right)^2 \left(\frac{r_0}{a}\right)^4
\left(\frac{B}{ B_0}\right)^2~W~~,
\end{equation}
where $B$  is the surface magnetic field of the pulsar, $B$ is assumed to be proportional to $M$.
According to (\ref{6.4b}) a null string passing a pulsar  in the near zone at a distance $a=10^2~km$, generates a EM pulse with the peak power
\begin{equation}\label{6.5}
\dot{E}\sim 10^{21}~\left(\frac{\omega}{\omega_0}\right)^2~W~~.
\end{equation}
For $\omega=\omega_0$ (\ref{6.5}) is still extremely small, if we compare it with the radiation power of the Sun,  $\dot{E}_\odot \sim10^{26}~W$,
or with the rate of rotation energy loss of the pulsar, $10^{30}~W$.  For a typical pulsar a null string may cause
considerable variations of its luminosity if $\omega \simeq 10^2 ~\omega_0$.

 Luminosity (\ref{6.4b})  can be increased if the string moves closer to the star or if the star has a larger magnetic moment. For instance, for a magnetar with the surface  magnetic field $B\sim 10^{15}~ G$ the estimate is
\begin{equation}\label{6.6}
\dot{E}\sim 10^{27}~\left(\frac{\omega}{ \omega_0}\right)^2~ W~~,
\end{equation}
which is comparable to the luminosity of a star even for $\omega \simeq \omega_0$.

Let us emphasize that the above estimates are quite rough. As has been shown in the previous Sections the intensity of the radiation
is not homogeneous and is maximal in the direction of the string velocity. In addition to the intensity and luminosity of the radiation the EM pulses generated by strings are specified by other physical parameters such as spectrum, duration of the pulse and etc.
The systems we consider, the string and the point sources, are characterized by an effective size of the interaction, which is the impact parameter $a$. This parameter determines the duration of the generated EM pulse and its typical frequency. For instance,
in the case of pulsars the duration of the pulse is $a/c\simeq 3\cdot 10^{-4}~s$.

It should be noted that for tensile cosmic strings there is a stronger limit on the energy, $G\mu \leq 10^{-12}$,
which is imposed by studying stochastic gravitational-wave background \cite{LIGOScientific:2021nrg}. If a similar restriction is applicable
to null strings the EM pulses generated by these strings are hardly observable.

\section{Discussion}\label{sum}
\setcounter{equation}0

The aim of this work was to find an analytical description of EM perturbations at future null infinity caused by null cosmic strings which move near charged particles or point  magnetic-dipole-like sources. By using the explicit asymptotic form of the perturbations we proved that
there is a non-vanishing flux of the radiation whose intensity is maximal in the direction of the velocity of the string. For strings
with energies $\omega\sim 10^{-5}$ moving in the near zones of pulsars the generated perturbations
of EM fields of the pulsars can cause considerable variations of the luminosities. For magnetars these variations can be caused by strings
with even lower energies, $\omega < 10^{-7}$.

We studied a simplified model of a straight null string with a constant energy per unit length. The string, however, changes its
form in the gravitational field of the source, and its energy density is not homegenously distributed over the length \cite{Davydov:2022qil}.
This effect should be taken into account in the subsequent calculations.

Although our computations have been done for point sources they can be easily extended to sources of a finite size. Then numerical methods can be used, say, to describe interaction of strings with more realistic fields of pulsars.

Another interesting development of our approach to asymptotic form of perturbations is application to perturbations of gravitational
fields of massive sources generated by null cosmic strings. By analogy with EM perturbations one may expect that null strings create
fluxes of gravitational waves. This work is in progress.

\section{Acknowledgments}

The authors are grateful to E.Davydov and V. Tainov for valuable discussions.
This research is supported by Russian Science Foundation grant No. 22-22-00684.

\bigskip
\bigskip
\bigskip

\newpage
\appendix

\section{Asymptotic for fields of electric type sources}\label{App1}
\setcounter{equation}0

Here we present details of estimate (\ref{5.7}) for $A_1$ in Sec. \ref{elrad}. Part $A_1$ of the vector potential,  see (\ref{5.10a}), is determined by (\ref{5.2})
when integration goes over a domain where $(\vec{m}\cdot \vec{n})+1$ is small.
To proceed with computations it is convenient to set coordinates $x^A={\theta,\varphi}$
\begin{equation}\label{a.5}
n_x=\cos\theta~~,~~n_y=\sin\theta\sin\varphi~~,~~n_z=\sin\theta\cos\varphi~~,
\end{equation}
where $n_i=x^i/r$, and use parametrization (\ref{5.52}) on the sphere
$S^2$  in (\ref{5.2}) with the help of the unit 3-vector $l^i$. It is convenient to change parametrization (\ref{5.52a}) as
\begin{equation}\label{a.7}
l_x=\cos\theta''~,~l_y=\sin\theta''\sin\varphi''~,~l_z=\sin\theta''
\cos\varphi''~~,
\end{equation}
to correspond to (\ref{a.5}).
In terms of $l$ the condition
\begin{equation}\label{a.8}
(\vec{m}\cdot \vec{n})=-1~~,
\end{equation}
which determines $A_1$, yields two solutions
\begin{equation}\label{a.9}
l^\pm_x=\pm\sqrt{\frac{1-n_x}{ 2}}~~,~~l^\pm_y=\mp \frac{n_y}{ \sqrt{2(1-n_x)}}~~,~~l^\pm_z=\mp \frac{n_z}{ \sqrt{2(1-n_x)}}~~,
\end{equation}
or $\theta''_\pm=(\pi\mp \theta)/2$,
$\varphi''_+=\varphi+\pi$, $\varphi''_-=\varphi$.  By using (\ref{a.7}), (\ref{a.9}) one gets the corresponding
quantities for these solutions:
\begin{equation}\label{a.10}
g(\Omega_+,\omega)=\frac{1}{ \sqrt{2(1-n_x)}}\left(-n_y+\omega (1-n_x)
+i\varepsilon^+\right)~~,~~g(\Omega_-,\omega)=-(g(\Omega_+,\omega))^*~~,
\end{equation}
\begin{equation}\label{a.11}
\varepsilon^{\pm}(\Omega_\pm)=\varepsilon_\pm=\pm\sqrt{2(1-n_x)-n_y^2}~~.
\end{equation}
It then follows that vector $\beta_\mu(\Omega)$ defined by (\ref{a.2}), (\ref{a.3}) has the property
\begin{equation}\label{a.14}
\beta^\mu(l^-)=(\beta^\mu(l^+))^*~~,
\end{equation}
where $\beta^\mu(l^\pm)\equiv \beta^\mu(\Omega_\pm)$.
Since we are estimating integral (\ref{5.2}) near $\theta''_\pm, \varphi''_\pm$ it is convenient
to introduce, near the each point new coordinates  $\zeta_1=\theta''-\theta''_\pm$, $\zeta_2=\varphi''-\varphi''_\pm$. At small $\zeta_A$
\begin{equation}\label{a.13}
(\vec{m}\cdot \vec{n})+1 \simeq 2\zeta_1^2+\frac 12 \sin^2\theta_\pm ~\zeta_2^2~~.
\end{equation}
The integration region is determined by the scale $\Lambda$, see (\ref{5.6}). If $|\zeta_A | < \Lambda$,
\begin{eqnarray}
A_{1,\mu} (x)&\equiv&-C \int_{|\zeta_A | < \Lambda} d\Omega''~\Re\left[\frac{\beta_\mu (\Omega'')}{ x^\nu ~m_\nu (\Omega'')+ia\varepsilon (\Omega'')}\right]
\nonumber \\
&\simeq& - \sum_{\pm} C\int_{|\zeta_A | < \Lambda}\sin\theta''_\pm d\zeta_1d\zeta_2~\Re\left[\frac{\beta_\mu (l^\pm)}{ U+r(2\zeta_1^2+\frac 12 \sin^2\theta~ \zeta_2^2)+ia\varepsilon^\pm}\right]
\nonumber\\
&=&-\frac{\pi C }{ r} \sum_{\pm} \frac{\sin\theta''_\pm }{ \sin \theta} \, \Re\left[\beta_\mu (l^\pm) \ln\left(\frac{U+ia\varepsilon^\pm+\Lambda^2 r }{ U+ia\varepsilon^\pm}\right)\right].
\label{a.15}
\end{eqnarray}
By taking into account (\ref{a.14}) and relation
$$
\frac{\sin\theta''_\pm }{ \sin \theta}=\frac{1}{ \sin (\theta/2)}=\sqrt{\frac{2}{ 1-n_x}}
$$
one gets
\begin{equation}\label{a.16}
A_{1,\mu} (x)\simeq \frac{N(\Omega)}{ r} ~
\Re\left[\bar{\beta}_\mu  \ln\left(\frac{U+ia\bar{\varepsilon}+\Lambda^2 r}{ U+ia\bar{\varepsilon}}\right)\right]~~,
\end{equation}
\begin{equation}\label{a.17}
\bar{\beta}_\mu= \beta_\mu (l^+)~~,~~\bar{\varepsilon}=\varepsilon^+~~,
\end{equation}
\begin{equation}\label{a.18}
N(\Omega)=-2\pi C \sqrt{\frac{2}{ 1-n_x}}~~.
\end{equation}
Eq. (\ref{a.16}) reproduces (\ref{5.7}).  The approximation is not applicable at $n_x=1$.

\section{Asymptotic for fields of magnetic type sources}\label{App2}
\setcounter{equation}0

Here we give some details of our computations in Sec. \ref{marad}. Consider solution to the following problem:
\begin{equation}\label{a2.17}
\Box \chi(u, {\bf x})=0~~,~~\chi(0, {\bf x})=\delta(y)~ f({\bf x})~~,
\end{equation}
which appears in the $y$-component in Eq. (\ref{5.30a}).
The solution can be written as
\begin{equation}\label{5.34}
\chi(u,v,y,z)=\int dy'dz'dv' D(u,v-v{\color{black}'},y-y',z-z')\delta(y')f(v',y',z')~~,
\end{equation}
where the $D$-function,
\begin{equation}\label{5.35}
D(x)=\frac{1}{ \pi}\frac{\partial}{ \partial v}\delta (x^2)~~,~~x^2=-uv+y^2+z^2~~,
\end{equation}
is the solution to the problem:
\begin{equation}\label{5.36}
\Box D(x)=0~~,~~D(u=0,{\bf x})=\delta^{(3)}({\bf x})~~,
\end{equation}
see details in \cite{Fursaev:2022ayo}. This yields $\chi=\partial_v\bar{\chi}$, where
\begin{equation}\label{a2.1} 
\bar{\chi}(u,v,y,z)=C \int^\infty_{-\infty} dz'\left[(uv-y^2{\color{black}-}(z-z')^2)^2{\color{black}}+4u^2((z')^2+a^2)\right]^{-1/2}~~,
\end{equation}
and $C=-1/(2\pi^2)$. In coordinates (\ref{4.5})
\begin{equation}\label{a2.2}
\bar{\chi}(x)=\frac{C}{r}X(U,r,\Omega)~~,~~X(U,r,\Omega)= \int^\infty_{-\infty} \frac{dx}{ f^{1/2}(x)}
\end{equation}
\begin{equation}\label{a2.3}
f(x)=({\color{black}-}x^2+2xn_z+c_1(2+c_1))^2+4(x^2+c_2^2)(c_1+(1-n_x))^2~~,
\end{equation}
where $x=z'/r$, $c_1={U/ r}$, $c_2={a/ r}$. To estimate $X$ at large $r$ we decompose $X=X_1+X_2$,
\begin{equation}\label{a2.4}
X_1(U,r,\Omega)= \int_{|x|<L} \frac{dx}{ f^{1/2}(x)}~~,~~X_2(U,r,\Omega)= \int_{|x|\geq L} \frac{dx}{ f^{1/2}(x)}~~.
\end{equation}
$L$ is some parameter which is assumed to be $L\gg c_i$. In the leading order
$X_2$ is a quasi-static function,
\begin{equation}\label{a2.5}
X_2(U,r,\Omega)\simeq  X_2(\Omega) = \int_{|x|\geq L}\frac{dx}{ x ((x{\color{black}-}2n_z)^2+4(1-n_x)^2)^{1/2}}~~.
\end{equation}
As for $X_1$, the main contribution to the integral comes from $x$ near the point $x_\star$, where $f(x)$ has a  minimum. At large $r$
\begin{equation}\label{a2.6}
f'(x_\star)=0~~,~~x_\star\simeq -\frac{n_zc_1}{ \bar{\varepsilon}^2}~~,
\end{equation}
\begin{equation}\label{a2.7}
f(x_\star)\simeq 4(1-n_x)^2~\frac{c_1 +\bar{\varepsilon}^2 c_2}{ \bar{\varepsilon}^2}~~,~~f''(x_\star)\simeq 8 \bar{\varepsilon}^2~~~.
\end{equation}
To guarantee that $f''(x_\star)>0$ we assume that $n_x\neq 1$. Then one has
$$
X_1(U,r,\Omega)\simeq  \int_{|x|<L} \frac{dx }{ \sqrt{f(x_\star)+f''(x_\star)(x-x_\star)^2/2}}=
$$
\begin{equation}\label{a2.8}
\frac{1}{ \sqrt{2f''(x_\star)}}\ln\left(\frac{\sqrt{\bar{L}^2+1} +\bar{L}}{
\sqrt{\bar{L}^2+1} - \bar{L}}\right)~~,~~ \bar{L}=L\sqrt{\frac{f''(x_\star) }{ 2|f(x_\star)|}}~~.
\end{equation}
This yields
\begin{equation}\label{a2.9}
X_1(U,r,\Omega)\simeq \frac{1}{ 4\bar{\varepsilon}}\ln \left(\frac{2Lr}{ (1-n_x)\sqrt{U^2 +a^2 \bar{\varepsilon}^2}} \right)~~,
\end{equation}
\begin{equation}\label{a2.10}
\bar{\chi}(x)=\frac{1}{ 2\pi^2 r}\left[\frac{1}{ 4\bar{\varepsilon}}\ln \left(\frac{\sqrt{U^2 +a^2 \bar{\varepsilon}^2}}{ 2Lr} \right)
-X(\Omega)\right]~~,
\end{equation}
where $X(\Omega)$ is some static part.

To find  asymptotic (\ref{5.43}) of $A_{2}(x)$ we compute the derivatives of $\bar{\chi}$
\begin{eqnarray}\label{a2.11}
\partial_y \bar{\chi}&=&n_y (\partial_r\bar{\chi}-\partial_U\bar{\chi})+\frac{1}{r}\frac{n_z}{1-n_x^2}
\partial_{\phi}\bar{\chi}+\frac{1}{r}\frac{n_x n_y}{\sqrt{1-n_x^2}}
\partial_{\theta}\bar{\chi}~,\\ \label{a2.12}
\partial_v \bar{\chi}
&=&
\frac{n_x}{ 2} \,\partial_r\bar{\chi}+\frac{({\color{black}1}-n_x)}{ 2}\partial_U\bar{\chi}{\color{black}-}\frac{\sqrt{1-n_x^2}}{2 r} \partial_{\theta} \bar{\chi}.
\end{eqnarray}
where
\begin{eqnarray}\label{a2.13}
\partial_r\bar{\chi}(x)&=&\frac{1}{ 2\pi^2 r^2}\left[X(\Omega)-\frac{1 }{ 4\bar{\varepsilon}}\left(\ln \left(\frac{\sqrt{U^2 +a^2 \bar{\varepsilon}^2}}{ 2Lr} \right)+1\right)\right],\\ \label{a2.14}
\partial_{b}\bar{\chi}(x)&=&\frac{1}{ 2\pi^2 r}\left[-\frac{1}{4 \bar{\varepsilon}^2}\ln \left(\frac{\sqrt{U^2 +a^2 \bar{\varepsilon}^2}}{ 2Lr} \right)\frac{\partial \bar{\varepsilon}}{\partial b}-\frac{a^2}{U^2+a^2\bar\varepsilon^2}\,\frac{\partial \bar{\varepsilon}}{\partial b}-\frac{\partial X}{\partial b}\right], \\
\partial_{U}\bar{\chi}(x)&=&\frac{1}{8\pi^2 r \,\bar{\varepsilon} }\, \frac{U}{U^2+a^2 \bar{\varepsilon}^2}.\label{a2.15}
\end{eqnarray}
with $b=\phi,\theta$. With the help of  (\ref{a2.11})-(\ref{a2.15}) we finally obtain
\begin{eqnarray}\label{a2.16}
&&A_{2,\nu}(x) ={\color{black}2}\omega M_z(2\delta_\nu^y \partial_v+\delta_\nu^u \partial_y)\bar{\chi}
\simeq {\color{black}\frac{\omega M_z}{\pi^2 r }\,\frac{(\delta_\nu^y {\color{black}(1-n_x)} -\delta_\nu^u n_y)}{4 \bar{\varepsilon}}} \frac{U}{U^2+a^2 \bar{\varepsilon}^2}.
\end{eqnarray}
The terms in the right hand side which decay as  $r^{-2}\ln r $ or faster are omitted.

\newpage
\bibliographystyle{unsrt}

\begin{thebibliography}{10}

\bibitem{Fursaev:2022ayo}  D.V. Fursaev, I.G. Pirozhenko, {\it Electrodynamics under the action of null cosmic strings}, Phys. Rev. {\bf D 107} (2023) no. 2, 025018, e-Print: 2212.05564 [gr-qc].

\bibitem{Schild:1976vq}
	A.~Schild.
	\newblock {Classical Null Strings}.
	\newblock {\em Phys. Rev. D}, 16:1722, 1977.
	
\bibitem{Fursaev:2021xlm}  D.V. Fursaev,  {\it Optical Equations for Null Strings}, Phys. Rev. {\bf D 103} (2021) no.12, 123526, e-Print: arXiv:2104.04982  [gr-qc].

\bibitem{Davydov:2022qil}  E.A. Davydov, D.V. Fursaev, V.A.Tainov,  {\it Null Cosmic Strings: Scattering by Black Holes, Optics, and Spacetime Content}, Phys. Rev. {\bf D 105} (2022) no.8, 083510, e-Print: arXiv:2203.02673 [gr-qc].


	
\bibitem{Kibble:1976sj}
	T.~W.~B. Kibble.
	\newblock {Topology of Cosmic Domains and Strings}.
	\newblock {\em J. Phys. A}, 9:1387--1398, 1976.
	
\bibitem{Vilenkin:2000jqa}
	A.~Vilenkin and E.~P.~S. Shellard.
	\newblock {\em {Cosmic Strings and Other Topological Defects}}.
	\newblock Cambridge University Press, 7 2000.


\bibitem{GM1} D. J. Gross and P. F. Mende, {\it The High-Energy Behavior of String Scattering
Amplitudes}, Phys. Lett. {\bf B 197} (1987) 129.

\bibitem{GM2} D. J. Gross and P. F. Mende, {\it String Theory Beyond the Planck Scale}, Nucl. Phys. {\bf B303} (1988) 407.

\bibitem{Lee:2019wij} Seung-Joo Lee, W. Lerche, T. Weigand, {\it Emergent Strings from Infinite Distance Limits},
e-Print: 1910.01135 [hep-th].

\bibitem{Xu:2020nlh} F. Xu, {\it On TCS $G_{2}$ manifolds and 4D emergent strings},
JHEP {\bf 10} (2020) 045, e-Print: 2006.02350 [hep-th].


\bibitem{Fursaev:2017aap}
	D.~V. Fursaev,
 {\it Physical Effects of Massless Cosmic Strings},
  Phys. Rev. {\bf D 96} (2017)  no. 10, 104005.
	
\bibitem{Fursaev:2018spa}
	D.~V. Fursaev,
	\newblock {\it Massless Cosmic Strings in Expanding Universe},
	\newblock {\em Phys. Rev. D}, 98(12):123531, 2018.


\bibitem{vandeMeent:2012gb}  M. van de Meent, {\it Geometry of Massless Cosmic Strings}, Phys. Rev. {\bf D87} (2013) no.2, 025020, e-Print: arXiv:1211.4365 [gr-qc].


\bibitem{Duval:2014lpa}   C. Duval, G.W. Gibbons, P.A. Horvathy, {\it Conformal Carroll groups}
J. Phys. {\bf A47} (2014) 33, 335204, e-Print: arXiv:1403.4213 [hep-th].
	
\bibitem{Duval:2014uoa}   C. Duval, G.W. Gibbons, P.A. Horvathy, P.M. Zhang
{\it Carroll versus Newton and Galilei: two dual non-Einsteinian concepts of time},
 Class. Quantum Grav. {\bf 31} (2014) 085016, e-Print: arXiv:1402.0657 [gr-qc].
 
\bibitem{Ciambelli:2023tzb}
L.~Ciambelli and D.~Grumiller,
{\it Carroll geodesics}, arXiv:2311.04112 [hep-th], 2023. 

\bibitem{Ciambelli:2023xqk}
L.~Ciambelli,
{\it Dynamics of Carrollian Scalar Fields}, arXiv:2311.04113 [hep-th], 2023.

\bibitem{Gourgoulhon:2005ng} E. Gourgoulhon and J.L. Jaramillo, {\it A 3+1 perspective on null hypersurfaces and isolated horizons}
Phys. Rept. {\bf 423} (2006) 159-294, e-Print: gr-qc/0503113 [gr-qc].

\bibitem{Penrose:1972xrn} R. Penrose, {\it The Geometry of Impulsive Gravitational Waves}, in
{\it General relativity: Papers in honour of J.L. Synge}, L. O'Raifeartaigh, ed. (1972), pp. 101--115.
	

\bibitem{Morse:1953} 	
P. M. Morse, H. Feshbach.
\newblock{\em Methods of Theoretical Physics},
\newblock{McGraw-Hill}, 1953, Part~I, ch. 6.1, page 683.





\bibitem{Landau2} L.D. Landau, E. M. Lifshitz, {\it The Classical Theory of Fields}, Vol. 2, Butterworth-Heinemann. ISBN 978-0-7506-2768-9  (1975).

\bibitem{Planck:2013mgr} P. A. R. Ade et al. [Planck], {\it Planck 2013 results. XXV. Searches for cosmic strings and other topological defects}, Astron. Astrophys. {\bf A25} (2014)  571, e-Print:1303.5085  [astro-ph.CO].

\bibitem{Charnock:2016nzm} T. Charnock, A. Avgoustidis, E. J. Copeland and
A. Moss, {\it CMB Constraints on Cosmic Strings and
Superstrings},  e-Print:1603.01275 [astro-ph.CO].

\bibitem{Dvorkin:2011aj} C. Dvorkin, M. Wyman and W. Hu, {\it Cosmic String
constraints from WMAP and the South Pole Telescope},
Phys. Rev. {\bf D 84} (2011) 123519, e-Print:1109.4947 [astro-ph.CO].

\bibitem{Beskin:2013UFN} V.S. Beskin, Ya.N. Istomin, A.A. Filippov,	{\it Radio pulsars: the search for truth},	Physics Uspekhi {\bf 56} (2) (2013).

\bibitem{Gunn:1969}  J.E. Gunn and  J.P. Ostriker,
{\it Acceleration of high-energy cosmic rays by pulsars}, Phys. Rev. Lett. {\bf 22} (1969) 729.


\bibitem{LIGOScientific:2021nrg}  The LIGO Scientific Collaboration, the Virgo Collaboration, the KAGRA Collaboration:  R. Abbott,  and others, {\it Constraints on Cosmic Strings Using Data from the Third Advanced LIGO-Virgo Observing Run}, Phys. Rev. Lett. {\bf 126} (2021) no.24,  241102, e-Print: arXiv:2101.12248  [gr-qc].
	
\end{thebibliography}

\end{document}